\documentclass{autart}
\usepackage{times,graphicx,latexsym,amsmath,amssymb,longtable,wrapfig,wasysym}
\usepackage[usenames]{color}
\usepackage[final]{showkeys}
\usepackage{datetime,enumitem,mathdots}
\usepackage{booktabs}
\usepackage{arydshln}
\usepackage{mathrsfs}

\usepackage{soul}

\renewcommand{\vec}{\mbox{\rm vec}}

\newtheorem{assumption}{Assumption}
\newtheorem{remark}{Remark}
\newtheorem{theorem}{Theorem}
\newtheorem{lemma}{Lemma}

\newif\ifsolutions
\solutionsfalse

\begin{document}
\begin{frontmatter}
\title{Subspace Identification with Multiple Data Records: unlocking the archive\thanksref{footnoteinfo}}
\thanks[footnoteinfo]{This paper was not presented at any IFAC 
meeting. Corresponding author C.~M.~Holcomb.}

\author[solar]{Chad M. Holcomb}\ead{holcomb\_chad\_m@solarturbines.com},
\author[ucsd]{Robert R. Bitmead}\ead{rbitmead@ucsd.edu},
\address[solar]{Solar Turbines Inc., 4200 Ruffin Road, San Diego CA, 92123-5398, USA.}
\address[ucsd]{Department of Mechanical \&\ Aerospace Engineering,
	University of California, San Diego,
	9500 Gilman Drive, La Jolla CA, 92093-0411, USA.}  
          

\begin{abstract}	
We develop an approach to subspace system identification using multiple data records and present a simple rank-based test for the adequacy of these data for fitting the unique linear, noise-free, dynamic model of prescribed state-vector, input-vector and output-vector dimensions. The approach is motivated by the prospect of sorting through archives of operational data and extracting a sequence of not-necessarily-contiguous data records individually insufficient for providing identifiability but collectively making this possible. The test of identifiability then becomes the sorting criterion for accepting or rejecting new data records. En passant, the familiar Hankel structure of the data matrices of subspace system identification is reinterpreted and revised.
\end{abstract}
\end{frontmatter}

\section{Introduction}

Dynamic System Identification is traditionally posed as the problem of fitting the parameters of a dynamic model of a system using a single contiguous record of input and output data. Because of the dynamics, breaking the record or using two non-contiguous parts is disjunctive to the fitting process, since it requires the treatment of additional initial conditions or the overwhelming of initial condition effects via large data sets. We shall demonstrate that segmented data need not stymie the identification. The vehicle for the analysis is subspace system identification (SSI) in general (and the MOESP algorithm in particular), which is amenable to multi-input/multi-output (MIMO) system identification and which is increasingly the platform of choice for many implementations of linear system identification.

The motivation for this problem stems from the advent of \textit{the cloud} and \textit{the industrial internet,} where the penetration of remotely accessible data logging capabilities and attendant expectations has expanded greatly. In contrast to this new capability and the associated desire to use this copious data for diagnostic and prognostic purposes lies the observation that in general operational data is too poorly excited or too infrequently excited to reveal the system dynamics reliably. Using a single very large data record swamps the informative data with the under-informative. This feature has been examined by Carrette et al. \cite{CarrettBastinGeninGevers_TSP:1996} with the conclusion that discarding uninformative data can lead to improved estimation accuracy. By contrast, if archival data can be used in pieces so as to yield collectively a well-excited \textit{experiment}, then the archive suddenly gains diagnostic interest and value. This is the driver for the new methodology and for the associated test of adequacy of cumulative excitation.

Peretzki et al. \cite{PeretzkiIsakssonBittencourtForsman_AIChE2011} provide a consistent analysis of the problems and values of archived industrial data for the extraction of parametric system models. They develop an approach to the assessment of signal data quality and determination of its adequacy for model fitting by performing a running condition number test on data segments. Their model structures consist of the output regressing on Laguerre filtered input signals. Thus, the models do not require initial states to accommodate non-contiguous segments. Their motivation and data analytical focus parallel our approach here, although we develop a subspace system identification algorithm and its attendant exact theory for identifiability using ideas from behavioral systems theory in comparison to their approach based on asymptotic normality.

Recently, other approaches to accommodate missing data in system identification have been advanced. Markovsky \cite{MarkovskyExactCDC:2013,MarkovskyApproxCDC:2013} considers small gaps (including periodic gaps) in a single data record. The approach is to subdivide the block-Hankel matrices of the single-record data equation into complete block-Hankel sub-matrices and then to use linear algebraic approaches to extract the underlying model. This yields a variant of SSI which focuses on preserving the model fitting aspects. However, the excitation questions are not broached. 


The paper has two distinct parts. The first develops the precise theory by dealing with multiple-segment system identification from noise-free data for exact linear systems of known degree. Signal identifiability conditions are presented in terms of rank. The second part presents an example from industrial gas turbine systems, where the rank-revealing singular value decomposition (SVD) replaces the rank calculations to accommodate noisy data and approximate linearity. 

Our approach is to present briefly the single-record SSI data equation and approach in Section~\ref{sec:sSSI} before posing the multi-record formulation in Section~\ref{sec:mSSI}, drawing on their similarity. This multiple-segment algorithm is an extension of that from \cite{ZhangBitmeadAutom:05,ZhangBitmeadTSP:2008} associated with radio channel equalization. The algorithm was presented in the form of this paper in \cite{HolcombBitmead_TurboExpo:2015} using the instrumental variable variant, but without the current theory. Theorem~\ref{th:youBeauty} is presented in Section~\ref{sec:yB} providing a testable sufficient identifiability condition on the data in this noise-free, exact-modeling environment. These results are a natural extension of those of Willems and Markovsky \cite{willems_persistency_2005,MarkovskyWillemsVanHuffelDeMoor:2006} and an improvement of the test of \cite{HolcombBitmead_TurboExpo:2015}, which did not present a proof. A simple computational example is given in this section to fix ideas. Then in Section~\ref{sec:turbo}, multiple archival gas turbine data sets are used to construct an informative \textit{experiment} from which a model can be calculated. These data sets are not contiguous. Nor is any one data set completely informative by itself. 

\section{Single-record subspace system identification\label{sec:sSSI}}
As indicated above, we focus on the MOESP algorithmic approach to SSI \cite{Verhaegen:94} for clarity. Since the underlying data equations, \eqref{eq:ssidata} and \eqref{eq:Ssidata}, are the same for all SSI approaches, such as CVA \cite{LarimoreCDC:1990}, N4SID \cite{VanOverscheeDeMoorAutom:1994,VanOverschee&DeMoor:96}, etc, the approach and data adequacy test are applicable.

\subsection{System}
Consider a linear, time-invariant, noise-free, discrete-time, MIMO, dynamic system $\mathcal{S}$, described by the state space system equations
\begin{align}\label{eq:ss}
\mathcal{S} := \left\lbrace \begin{array}{rl}
x_{t+1} &= A x_{t} + B u_{t},\\
y_{t} &= C x_{t} + D u_{t},
\end{array}\right.
\end{align}
with input $u_{t} \in \mathbb{R}^{m}$, output $y_{t} \in \mathbb{R}^{p}$, and state $x_{t} \in \mathbb{R}^{n}$. 
\begin{assumption}\label{ass:sass}
For system $\mathcal S$ in \eqref{eq:ss} the dimensions $n$, $m$, $p$ are known and $n$ is the minimal state dimension. That is, $\mathcal S$ is observable and controllable. Signals $\{u_t,y_t:t=1,\dots,N\}$ are measured exactly.
\end{assumption}
Denote the following matrices.
\begin{align}
U_\ell&=\begin{bmatrix}
u_1&u_2&u_3&\dots&u_j\\
u_2&u_3&u_4&\iddots&u_{j+1}\\
u_3&u_4&\iddots&&u_{j+2}\\
\vdots&\iddots&&&\vdots\\
u_\ell&u_{\ell+1}&u_{\ell+2}&\dots&u_{\ell+j-1}
\end{bmatrix},\label{eq:Ul}
\end{align}
\begin{align}
Y_\ell&=\begin{bmatrix}
y_1&y_2&y_3&\dots&y_j\\
y_2&y_3&y_4&\iddots&y_{j+1}\\
y_3&y_4&\iddots&&y_{j+2}\\
\vdots&\iddots&&&\vdots\\
y_\ell&y_{\ell+1}&y_{\ell+2}&\dots&y_{\ell+j-1}
\end{bmatrix},\label{eq:Yl}
\end{align}

\begin{align}
\Gamma_{\ell} &=	\begin{bmatrix}
C \\ CA \\ \vdots \\ CA^{\ell-1}
\end{bmatrix},\label{eq:obsv}
\end{align}
\begin{align}
H_{\ell} &= \begin{bmatrix} 
D & 0  &	0&0&\cdots 	& 0 \\
CB   &  D   & 0 	&0&	\cdots 	& 0 \\
CAB 	& CB & D & 0 &\cdots 	& 0 \\
\vdots&\vdots & \vdots &	\ddots&&\vdots \\
CA^{\ell-2}B &	&&& 	\cdots  & D
\end{bmatrix},\label{eq:H}
\end{align}
\begin{align}
X_\ell &= \begin{bmatrix} x_1& x_2& \cdots & x_j \end{bmatrix}.\label{eq:Xl}
\end{align}
Their dimensions are $\ell m\times j$, $\ell p\times j$, $\ell p\times n$, $\ell p\times\ell m$, and $n\times j$, respectively. Note that: $U_\ell$ and $Y_\ell$ are block-Hankel data matrices, $\Gamma_\ell$ is the extended observability matrix, $H_\ell$ is a block-Toeplitz matrix of impulse response parameters, and $X_\ell$ is a matrix of successive state values.

The noise-free data equation of subspace system identification using a single segment is as follows.
\begin{align}\label{eq:ssidata}
Y_\ell&=\Gamma_\ell X_\ell+H_\ell U_\ell,\\
\begin{bmatrix}U_\ell\\Y_\ell\end{bmatrix}
&=\begin{bmatrix}I&0\\H_\ell&\Gamma_\ell\end{bmatrix}
\begin{bmatrix}U_\ell\\X_\ell\end{bmatrix}.\label{eq:blocker}
\end{align}

Subspace system identification proceeds in four steps from \eqref{eq:ssidata}.

\textbf{Step S.i:} Orthogonal projection to $U_\ell.$ Define
\begin{align}\label{eq:piperp}
\Pi_{U_{\ell}}^{\perp} = I -  U_{\ell}^T (U_{\ell} U_{\ell}^T)^{\dag}U_{\ell},
\end{align}
and multiply the data equation \eqref{eq:ssidata} on the right to yield
\begin{align}
Y_\ell \Pi^{\perp}_{U_\ell} =&
\Gamma_{\ell} X_\ell \Pi^{\perp}_{U_\ell} \label{eq:yperp}.
\end{align}
The column space of $\Gamma_\ell$ is the column space of $Y_\ell \Pi^{\perp}_{U_\ell} $.

\textbf{Step S.ii:} Extract the column space of $\Gamma_\ell$ using the singular value decomposition.
\begin{align}\label{eq:svd}
Y_\ell \Pi^{\perp}_{U_\ell}&= \begin{bmatrix} \mathcal{Q}_1 & \mathcal{Q}_2 \end{bmatrix} 
\begin{bmatrix} \Sigma_1 & 0 \\ 0 & \Sigma_2 \end{bmatrix} 
\begin{bmatrix} \mathcal{P}_1^T \\ \mathcal{P}_2^T \end{bmatrix},\\
\label{eq:gammah}
\hat\Gamma_\ell &= \mathcal{Q}_1\Sigma_1^{1/2}.
\end{align}
{}[Note that weighting matrices can be included in this step which specifies which variant of subspace system identification, N4SID, MOESP, CVA, etc, is being used. This is immaterial to our analysis here.]

\textbf{Step S.iii:} Take the first $p$ rows of $\hat\Gamma_\ell$ to be $\hat C$.
\begin{align}
\hat{C} = \hat{\Gamma}_l(1:p,:)\label{eq:Ch},
\end{align}
and solve  a least-squares problem for $\hat A$.
\begin{align}
\hat A&=\left(\hat\Gamma_\ell(1:(\ell-1)p,:)\right)^\dagger\hat\Gamma(p+1:\ell p,:).\label{eq:Ah}
\end{align}

\textbf{Step S.iv:} Solve a least-squares problem for $\hat B,$ $\hat D,$ and $\hat x_1$ as follows.
Write
\begin{align}
y_{t}=& 
\sum\limits_{\tau=1}^{t-1} CA^{t-1-\tau} B u_{\tau} + Du_t 
+ CA^{t-1} x_1.\label{eq:yseg}
\end{align}
Apply the vec operator and the Kronecker product $(\otimes)$ formula to \eqref{eq:yseg} to yield
\begin{align}\label{eq:vecls}
y_t&=\vec(y_t),\nonumber\\
 &= \phi_{B}^T(t)\vec(B)  + \phi_{D}^T(t)\vec(D)+\phi_{x_1}^T(t)x_1,
\end{align}
where
\begin{align}
\phi_{B}^T(t) &= \sum_{\tau=1}^{t-1}  
u_\tau^T \otimes \hat C\hat A^{t-1-\tau}, \label{eq:phib} \\
\phi_{D}^T(t) &= u_t^T \otimes I_{p}  \label{eq:phid},\\
\phi_{x_1}^T(t) &= \hat C\hat A^{t-1}  \label{eq:phixo}.
\end{align}
Now assemble the output and regressor matrices as follows.
\begin{align}
\mathsf Y&=\begin{bmatrix}y_1^T&y_2^T&\dots&y_N^T\end{bmatrix}^T,\label{eq:Y}\\
\Phi^T &=\begin{bmatrix} \Phi_B^T& \Phi_D^T& \Phi_{x_1}^T \end{bmatrix},\label{eq:PHI}
\end{align}
with submatrices
\begin{align}
\Phi_B &= \begin{bmatrix} \phi_{B}(1) & \cdots & \phi_{B}(N) \end{bmatrix},\\
\Phi_D&=\begin{bmatrix} \phi_{D}(1)	& \cdots & \phi_{D}(N) \end{bmatrix}, \\
\Phi_{x_1}&= \begin{bmatrix} \phi_{x_1}(1) & \cdots & \phi_{x_1}(N)\end{bmatrix}.
\end{align}
Define the parameter $(nm+mp+n)$-vector
\begin{align}
\theta &=\begin{bmatrix}\vec(B)^T & \vec(D)^T & x_1^T\end{bmatrix}^T.\label{eq:theta}
\end{align}
The least-squares problem for $B$, $D$, and $x_1$ is
\begin{align*}
\mathsf Y&=\Phi^T\theta,
\end{align*}
with estimate
\begin{align*}
\hat{\theta} &=\begin{bmatrix}\vec(\hat B)\\\vec(\hat D)\\\hat x_1\end{bmatrix}
= (\Phi\Phi^T)^{-1}\,\Phi \mathsf Y.
\end{align*}

\section{Multi-record subspace system identification\label{sec:mSSI}}
In place of the single data record $\{u_t,y_t:1=1,2,\dots,N\},$ of which we use $\ell+j$ elements, suppose we have available $j$ data records of length $\ell$: $$\bigg\{\{u_t,y_t:t=t_i+1,2,\dots,t_i+\ell\}:i=1,2,\dots,j\bigg\}.$$ Each record is a contiguous length-$\ell$ sequence of input and output signal values from System $\mathcal S$. Each of the $j$ records has unknown initial state vector $x_{t_i+1}$.

Define the following matrices
\begin{align}
\mathbb U_\ell&=\begin{bmatrix}
u_{t_1+1}&u_{t_2+1}&u_{t_3+1}&\dots&u_{t_j+1}\\
u_{t_1+2}&u_{t_2+2}&u_{t_3+2}&\dots&u_{t_j+2}\\
u_{t_1+3}&u_{t_2+3}&u_{t_3+3}&\dots&u_{t_j+3}\\
\vdots&\vdots&\vdots&&\vdots\\
u_{t_1+\ell}&u_{t_2+\ell}&u_{t_3+\ell}&\dots&u_{t_j+\ell}
\end{bmatrix},\label{eq:UUl}\\
\mathbb Y_\ell&=\begin{bmatrix}
y_{t_1+1}&y_{t_2+1}&y_{t_3+1}&\dots&y_{t_j+1}\\
y_{t_1+2}&y_{t_2+2}&y_{t_3+2}&\dots&y_{t_j+2}\\
y_{t_1+3}&y_{t_2+3}&y_{t_3+3}&\dots&_{t_j+3}\\
\vdots&\vdots&\vdots&&\vdots\\
y_{t_1+\ell}&y_{t_2+\ell}&y_{t_3+\ell}&\dots&y_{t_j+\ell}
\end{bmatrix},\label{eq:YYl}\\
\mathbb X_\ell &= \begin{bmatrix} x_{t_1+1}& x_{t_2+1}& \cdots & x_{t_j+1} \end{bmatrix}.\label{eq:XXl}
\end{align}
Then a variant of \eqref{eq:ssidata} holds with the same $\Gamma_\ell$ and $H_\ell$. To wit,
\begin{align}\label{eq:Ssidata}
\mathbb Y_\ell&=\Gamma_\ell \mathbb X_\ell+H_\ell \mathbb U_\ell,\\
\begin{bmatrix}\mathbb U_\ell\\\mathbb Y_\ell\end{bmatrix}
&=\begin{bmatrix}I&0\\H_\ell&\Gamma_\ell\end{bmatrix}
\begin{bmatrix}\mathbb U_\ell\\\mathbb X_\ell\end{bmatrix}.\label{eq:Blocker}
\end{align}
The central distinction between the single-record \eqref{eq:ssidata} and multi-record \eqref{eq:Ssidata} lies in the non-block-Hankel structure of the data matrices $\mathbb U_\ell$ and $\mathbb Y_\ell$. Although, one could return to \eqref{eq:ssidata} exactly from \eqref{eq:Ssidata} by selecting data records with 
\begin{align*}
t_1=0,\;t_2=1,\;\dots,t_j=j-1.
\end{align*}
So the single-record analysis can be subsumed in the multi-record approach.

The solution algorithm for recovering $\hat C$ and $\hat A$ proceeds precisely as in Steps \textbf{S.(i-iii)} of the single-record algorithm:

\textbf{Step M.i:} Multiply $\mathbb Y_\ell$ on the right by
\begin{align}\label{eq:Piperp}
\Pi_{\mathbb U_{\ell}}^{\perp} = I - \mathbb U_{\ell}^T (\mathbb U_{\ell} \mathbb U_{\ell}^T)^{\dag}\mathbb U_{\ell}.
\end{align}

\textbf{Step M.ii} Compute the SVD 
\begin{align*}
\mathbb Y_\ell\Pi_{\mathbb U_{\ell}}^{\perp}=
\begin{bmatrix}\mathbb Q_1&\mathbb Q_2\end{bmatrix}
\begin{bmatrix}\mathbb S_1&0\\0&\mathbb S_2\end{bmatrix}
\begin{bmatrix}\mathbb P_1^T\\\mathbb P_2^T\end{bmatrix}
\end{align*}
and take 
\begin{align*}
\hat \Gamma_\ell&=\mathbb Q_1\mathbb S_1^{1/2}.
\end{align*}

\textbf{Step M.iii:} Extract $\hat C$ and $\hat A$ as earlier from $\hat\Gamma_\ell$ via \eqref{eq:Ch} and \eqref{eq:Ah}.

\textbf{Step M.iv:} Rewrite \eqref{eq:yseg} for each record, $t=1,\dots,\ell,$ and record number, $i=1,\dots,j,$
\begin{align}
y_{t_i+t}=& 
\sum\limits_{\tau=1}^{t-1} \hat C\hat A^{t-1-\tau} B u_{t_i+\tau} + Du_{t_i+t} 
+ \hat C\hat A^{t-1} x_{t_i+1}.\label{eq:Yseg}
\end{align}
Whence, using (\ref{eq:phib}-\ref{eq:phixo}),
\begin{align}\label{eq:vecls1}
y_{t_i+t}
 &= \phi_{B}^T(t_i+t)\vec(B)\nonumber  + \phi_{D}^T(t_i+t)\vec(D)\\&\hskip 28mm +\phi_{x_{t_i+1}}^T(t)x_{t_i+1},
\end{align}
where we have deliberately indicated that the regressor terms for the initial conditions depend only on $t$ and not on $t_i$.

Define an extended parameter $(nm+mp+nj)$-vector
\begin{align}\label{eq:varthetadefn}
\vartheta&=\begin{bmatrix}\vec(B)^T&\vec(D)^T&x_{t_1+1}^T&x_{t_2+1}^T&\dots&x^T_{t_j+1}\end{bmatrix}^T.
\end{align}
Then the least-squares problem for $B$, $D$ and the initial conditions has the form
\begin{align}
\vec(\mathbb Y_\ell)=\Upsilon\vartheta,
\end{align}
with regression matrix, $\Upsilon$, of the structure
\begin{align}
\begin{bmatrix}
\Phi_B^T(t_1+1)&\Phi_D^T(t_1+1)&\hat C&0&\dots&0\\
\Phi_B^T(t_1+2)&\Phi_D^T(t_1+2)&\hat C\hat A&0&\dots&0\\
\vdots&\vdots&\vdots&\vdots&&\vdots\\
\Phi_B^T(t_1+\ell)&\Phi_D^T(t_1+\ell)&\hat C\hat A^{\ell-1}&0&\dots&0\\
\Phi_B^T(t_2+1)&\Phi_D(t_2+1)&0&\hat C&\dots&0\\
\vdots&\vdots&\vdots&\vdots&&\vdots\\
\Phi_B^T(t_2+\ell)&\Phi_D^T(t_2+\ell)&0&\hat C\hat A^{\ell-1}&\dots&0\\
\vdots&\vdots&\vdots&\vdots&&\vdots\\
\Phi_B^T(t_j+1)&\Phi_D(t_j+1)&0&0&\dots&\hat C\\
\vdots&\vdots&\vdots&\vdots&&\vdots\\
\Phi_B^T(t_j+\ell)&\Phi_D^T(t_j+\ell)&0&0&\dots&\hat C\hat A^{\ell-1}
\end{bmatrix}\label{eq:Upsilon}
\end{align}
Then the least-squares estimate
\begin{align}\label{eq:thetasoln}
\hat\vartheta&=(\Upsilon^T\Upsilon)^\dag\Upsilon^T\vec(\mathbb Y_\ell),
\end{align}
and $B$, $D$, $\{x_{t_i+1}: i=1,\dots,j\}$ are unpacked from $\hat\vartheta$ using \eqref{eq:varthetadefn}.

\section{Identifiability from multiple data records\label{sec:yB}}
Our main result follows. Its proof is given in Section~\ref{sec:theory} as is the definition of the \textit{maximal system lag,} $L$, which is overbounded by the system state dimension, $n$.
\begin{theorem}\label{th:youBeauty}
For 
$\ell> L,$ system $\mathcal S$'s maximal lag, the multiple record subspace system identification algorithm applied to data satisfying Assumption~\ref{ass:sass} yields a minimal state-space description of System $\mathcal S$ and all of its corresponding initial state values, $\{x_{t_1+1},\dots,x_{t_j+1}\}$ if the following two rank conditions are met.
\begin{align}\label{eq:Wrank}
\text{\rm rank}\,\begin{bmatrix}\mathbb U_\ell\\\mathbb Y_\ell\end{bmatrix}&=m\ell+n,
\end{align}
and
\begin{align}\label{eq:Urank}
\text{\rm rank}\,\mathbb U_\ell&=m\ell.
\end{align}
\end{theorem}
We note that an immediate requirement of Theorem~\ref{th:youBeauty} is that the data matrices $\mathbb U_\ell$ and $\mathbb Y_\ell$ contain sufficient columns,
\begin{align}\label{eq:colcount}
j\geq m\ell+n,
\end{align}
that \eqref{eq:Wrank} and \eqref{eq:Urank} might feasibly be satisfied. That is, at least $m\ell+n$ contiguous length-$\ell$ data records are required.

\subsection{Noise-free computational example}
To demonstrate the method's capabilities with noise-free data and accurate computation, we briefly present a computational example using MATLAB. We select the discrete-time, single-input/single-output, noise-free, minimal, second-order, linear time-invariant system with matrices
\begin{align*}
A=\begin{bmatrix}0.9&0.2\\0&0.8\end{bmatrix},\hskip 5mm
B=\begin{bmatrix}1\\1\end{bmatrix},\hskip 5mm
C=\begin{bmatrix}1&1\end{bmatrix},\hskip 5mm D=1.
\end{align*}
Here
\begin{align*}
m=1,\hskip 5mm n=2,\hskip 5mm p=1.
\end{align*}
We select data matrix block dimension 
\begin{align*}
\ell=3>L=n=2.
\end{align*}
{}[We note that for a proper single-input/single-output system such as this, the system maximal lag, $L,$ (developed in Section ~\ref{sec:theory}) equals the state dimension, $n$, and thus $\ell=3$ is the minimal possible choice.]

Seven data sets are constructed from specified initial states, running for 20 time steps, with input signals each drawn from independent, uniform $[0,1]$ white noise processes. Data record selection vector
\begin{align*}
\texttt{jayVec}=\begin{bmatrix}0&0&0&2&2&1&0\end{bmatrix},
\end{align*}
dictates that two length-3 data columns are drawn from Records~4 and 5, i.e. the first four data from each, while one length-3 column is drawn from Record~6. The number of $(u_t,y_t)$ data pairs used is 11 and the number of columns is
\begin{align*}
j=5=m\ell+n.
\end{align*}
The initial states for these three records are chosen to be
\begin{align*}
x_{4,1}=\begin{bmatrix}-1\\-1\end{bmatrix},\hskip 5mm
x_{5,1}=\begin{bmatrix}0.5\\1\end{bmatrix},\hskip 5mm
x_{6,1}=\begin{bmatrix}1\\0.5\end{bmatrix}.\hskip 5mm
\end{align*}

The data matrix $\mathbb U_\ell$ is $3\times 5,$ has singular values $(2.0142, 0.2645, 0.0906)$ and therefore is rank $m\ell=3$. Condition \eqref{eq:Urank} holds. Further, using svd to compute rank,
\begin{align*}
\texttt{svd}\begin{bmatrix}\mathbb U_\ell\\\mathbb Y_\ell\end{bmatrix}=
(10.6335, 2.2668, 0.2509, 0.0998, 0.0221),
\end{align*}
and condition \eqref{eq:Wrank}, rank=5, holds for this data selection.

The identified state space realization matrices and initial states for each record are 
\begin{align*}
\hat C&=\begin{bmatrix}-0.6084&0.7157\end{bmatrix},\hskip 5mm
\hat A=\begin{bmatrix}0.9428&1.0603\\-0.0058&0.7572\end{bmatrix},\\
\hat B&=\begin{bmatrix}-3.2885\\-0.0009\end{bmatrix},\hskip 5mm
\hat D=1.0000,\\
\hat x_{4,1}&=\begin{bmatrix}3.2885\\0.0009  \end{bmatrix},
\hat x_{5,1}=\begin{bmatrix}-2.5039\\-0.0326  \end{bmatrix},
\hat x_{6,1}=\begin{bmatrix}-2.4288\\0.0312  \end{bmatrix}.
\end{align*}
We recover the state transformation matrix
\begin{align*}
T=\begin{bmatrix}-0.3064&8.1890\\-0.3020&-7.4733\end{bmatrix},
\end{align*}
from which it is simply demonstrated that 
\begin{align*}
&A=T\hat AT^{-1},\hskip 3mm B=T\hat B,\hskip 3mm C=\hat CT^{-1},\hskip 3mm \hat D=D,&\\
&\begin{bmatrix}x_{4,1}&x_{5,1}&x_{6,1}\end{bmatrix}=
T\begin{bmatrix}\hat x_{4,1}&\hat x_{5,1}&\hat x_{6,1}\end{bmatrix}.&
\end{align*}
Regression matrix, $\Upsilon,$ of \eqref{eq:Upsilon} is $11\times 9$ with singular values (3.0502, 1.7359, 1.3986, 1.1391, 1.0001, 1.0000, 0.6407, 0.2578, 0.0635). The 11 rows correspond to the data $\{y_{4,1},\,y_{4,2},\, y_{4,3},\, y_{4,4},\, y_{5,1},\, y_{5,2},\, y_{5,3},\, y_{5,4},\, y_{6,1},\, y_{6,2},\, y_{6,3}\}$. The 9 columns comprise 2 regressors for the elements of $\hat B$, one regressor for $\hat D$, and 6 regressors for the elements of $\hat x_{4,1},$ $\hat x_{5,1},$ and $\hat x_{6,1}$.

\section{Theory\label{sec:theory}}

We make the following definitions.
\begin{description}
\item[Matrix $W_{\ell,j}$:] For given $\ell$ and $j$, 
\begin{align}
\hskip -5mm w_t=\begin{bmatrix}u_t\\y_t\end{bmatrix}
\;\;\text{and}\;\;\;W_{\ell,j}&=\begin{bmatrix}
w_1&w_2&\dots&w_j\\
w_2&w_3&\iddots&w_{j+1}\\
\vdots&\iddots&&\vdots\\
w_\ell&w_{\ell+1}&\dots&w_{\ell+j-1}
\end{bmatrix}.\label{eq:Wl}
\end{align}
\item[Behavior of $\mathcal S$:] The set of all signal functions $\{w_t; t=1,\dots\}$ satisfying $\mathcal S$ is the behavior of $\mathcal S$, denoted $\mathfrak B$. The length-$\ell$ behavior of system $\mathcal S$, denoted $\mathfrak B_{[1,\ell]},$ is the span of the column space of $W_{\ell,j}$ as initial states and signals $\{w_t\}$ vary over their entire range for all signals satisfying \eqref{eq:ss}.
\item[Annihilator Polynomial Matrix $R(\sigma)$:] Minimal-degree, row-reduced $p\times (p+m)$ matrix polynomial $R(\sigma)$ in backward shift operator $\sigma$ (unique up to left multiplication by a unimodular polynomial matrix) satisfying
\begin{align}\label{eq:reqn}
R(\sigma)w_t=R^\prime(\sigma)\xi_t,
\end{align}
for all $w_t\in\mathfrak B$ and for auxiliary variables (states here) $\xi_t$ and auxiliary $p\times n$ polynomial matrix $R^\prime,$ is the minimal annihilator polynomial of the behavior. Such $R(\sigma)$ yields the left Matrix Fraction Description of the transfer function matrix of $\mathcal S$ and, hence, uniquely identifies $\mathcal S$. This is known as the \textit{kernel representation of $\mathcal S$}. Thereom~1 of \cite{Willems_Autom:86} establishes that the presence of the states does not affect the choice of annihilator $R$.
\item[Maximal lag $L$:] The maximal degree of the elements of $R(\sigma)$ is the maximal system lag, $L$. Polynomial matrix $R(\sigma)$ has $L+1$ constant $p\times (p+m)$ matrix coefficients. Evidently, $L\leq n,$ the system degree and minimal state dimension.
\end{description}
We have the following property characterizing the dimension of the range of $\mathfrak B_{[1,\ell]}$.
\begin{lemma}[Willems (1986)]\label{lem:willdim}
For $\ell>L$ and minimal $\mathcal S$,
\begin{align}\label{eq:dimcond}
{\rm dim}\,\mathfrak B_{[1,\ell]}=m\ell+n.
\end{align}
\end{lemma}
Clearly the left nullspace of $W_{\ell,j}\supseteq{\rm ker }\;\mathfrak B_{[1,\ell]}$ and, if rank~$W_\ell={\rm dim}\;\mathfrak B_{[1,\ell]},$ then left nullspace $W_{\ell,j}={\rm ker }\;\mathfrak B_{[1,\ell]}$ and the kernel representation of $\mathcal S$ is recovered from this left nullspace. Whence, 
\begin{lemma}[Willems et al (2005)]\label{lem:will2}
For $\ell> L$ and minimal $\mathcal S$ if,
\begin{align}\label{eq:niceW}
{\rm rank}\,W_{\ell,j}=m\ell+n,
\end{align}
then the left nullspace of $W_{\ell,j}$ uniquely identifies system $\mathcal S$.
\end{lemma}

The precise condition, $\ell>L$, which in \cite{Willems_Autom:86,MarkovskyWillemsVanHuffelDeMoor:2006} is difficult to discern because of somewhat mercurial notation, follows from the isomorphism between the left nullspace of $W_{\ell,j}$ and the $\ell$ coefficients of $R(\sigma)$. Since the degree of $R$ is $L$, there is requirement for at least $L+1$ coefficient matrices. So $\ell$ must equal or exceed $L+1.$

\begin{lemma}\label{lem:wrank}
For $\ell> L$ and minimal $\mathcal S$, if \eqref{eq:Wrank},
\begin{align*}
{\rm rank}\,\begin{bmatrix}\mathbb U_\ell\\\mathbb Y_\ell\end{bmatrix}&=m\ell+n,
\end{align*}
holds then the left nullspace of  the multi-record data matrix $\begin{bmatrix}\mathbb U_\ell\\\mathbb Y_\ell\end{bmatrix}$ uniquely identifies $\mathcal S$.
\end{lemma}
\textit{\underline{Proof}:} The data matrix from \eqref{eq:blocker} is a row permutation of the corresponding matrix $W_{\ell,j}$. Thus the ranks and left nullities of the two matrices are the same. Further, the columns of the multiple-record data matrix of \eqref{eq:Wrank} are columns of $W_{\ell,j}$ for $j=t_j+\ell$ and this subset of columns spans $\mathfrak B_{[1,\ell]}$ if condition \eqref{eq:Wrank} holds. Thus, the kernel representation of $\mathcal S$ will then be recoverable from the left nullspace of the multiple data matrix and the system is identifiable from this data.
\hfill$\square$

\begin{lemma}\label{lem:ics}
If minimal $\mathcal S$ is uniquely identifiable and $\ell\geq L$, then the initial state for each column of 
$\begin{bmatrix}\mathbb U_\ell\\\mathbb Y_\ell\end{bmatrix}$ is uniquely calculable from that column.
\end{lemma}
\textit{\underline{Proof}:} This follows from the observability of $\mathcal S$ and the connection between the maximal lag, $L$, and the observability indices \cite{Willems_Autom:86,Wolovich:1974}.\hfill$\square$

\subsection*{Proof of Theorem~\ref{th:youBeauty}}
From Lemma~\ref{lem:wrank}, \eqref{eq:Wrank} suffices for unique identifiability of $\mathcal S$ via the nullspace of the multi-record data matrix. Condition \eqref{eq:Urank} further ensures that, within the data records, we are free to choose the $\{u_t\}$ as the input signal. In the behavioral approach, Willems \cite{Willems_Autom:86} defines the input signals to be those elements which may be taken as free variables in the data; a property associated with \eqref{eq:Urank}. More particularly, this condition assures that $\Pi_{\mathbb U_{\ell}}^{\perp}$ of \eqref{eq:Piperp} has rank precisely $n$, thereby permitting the exact recovery of the rank $n$ column space of the extended observability matrix.\hfill$\square$

\begin{remark}
We note that condition $\ell>L$ in Theorem~\ref{th:youBeauty} might be replaced by $\ell>n,$ the minimal state dimension, in case $L$ is unknown, which is frequently the case in system identification. For SISO systems, $L=n$. But for MIMO systems, $L\leq n$. 
\end{remark}
\begin{remark}
It is instructive to compare the persistence of excitation condition derived in \cite{WillemsRapisardaMarkovskyDeMoorSCL:2005} to \eqref{eq:Urank}. Willems et al. appeal to \eqref{eq:niceW} in Lemma~\ref{lem:will2} for identifiability and then seek a condition on the single-record $\{u_t: t=1,\dots,\ell+j-1\}$ sequence alone which is capable of ensuring \eqref{eq:niceW}. His condition is that 
\begin{align}
{\rm rank}\;U_{\ell+n}=m(\ell+n),\label{eq:willcond}
\end{align}
where $U_\ell$ is defined in \eqref{eq:Ul}. Interestingly, this latter condition requires satisfaction of a rank property on signals, part of which do not appear in the data set used for identification. One suspects that for quasi-stationary data there is a frequency content interpretation.

From the perspective of multiple data records, a surprising alteration to the theory is the disappearance of block-Hankel matrices. Single contiguous length-$N$ data records yield $N-\ell+1$ candidate length-$\ell$ contiguous sub-records which might be stacked into block-Hankel matrices -- their column order is unimportant to the problem. Yet Hankel matrices are fundamental to much of Systems Theory and realizations. Property \eqref{eq:willcond} preserves this focus on Hankel.
\end{remark}

\begin{remark}
It is clear that the linear algebraic approach to the multi-record problem parallels that of the single-record method with the inherent inclusion of more initial conditions and a corresponding complication of the regression problem. However, it is also evident from Lemma~\ref{lem:ics} that the estimation of these additional initial states is no more problematic than the estimation of a single initial state, since the corresponding regression matrix is a direct sum of extended observability matrices, $\hat \Gamma_\ell$, which are full column rank by construction. This mirrors the content of Thereom~1 of \cite{Willems_Autom:86} in showing that the kernel estimation problem is unencumbered by the presence of initial states. In practice, the incorporation of additional initial conditions can introduce numerical issues.
\end{remark}

\begin{remark}[The {[low]} price of multiple data segments]
For the noise-free exact computation of Section~\ref{sec:yB}, the data segment selection matrix \verb!jayVec! was introduced to choose the number of columns of the data matrix to be composed from each available segment -- the example chooses two columns from segments four and five plus a single column from segment six, which involved precisely eleven input-output data pairs. 

The example uses state dimension $n=2$ and minimal $\ell=3$ and minimal $j=m\ell=5.$ Its second-order scalar transfer function with direct feedthrough,
\begin{align*}
P(z)&=\frac{b_0+b_1z^{-1}+b_2z^{-2}}{1+a_1z^{-1}+a_2z^{-2}},
\end{align*}
possesses exactly five independent parameters. By the same token, each initial state requires two independent parameters.

Examining the exact minimal number of input-output data pairs implied by (\ref{eq:UUl}-\ref{eq:YYl}) for this selection of $n$, $\ell$ and $j$, we see that: using a single data segment requires exactly seven data; using two data segments requires exactly nine data; three data segments requires eleven data, etc. So the additional initial conditions implicitly force more data pairs to be used in the construction of the data matrices, even though the data matrix dimension does not alter.
\end{remark}

\section{Archival industrial gas turbine data sets}\label{sec:turbo}


Part~2 of this paper applies these theoretically supported exact results based on rank conditions to industrial data sets in order to demonstrate several features:
\begin{enumerate}[label=(\roman*)]
\item that the methods are amenable to handling less than pristine industrial data, where linearity is only a local feature, models are admittedly approximate, and data integrity is challenged;
\item that the rank conditions of the theory can be effectively replaced by rank-revealing SVD conditions;
\item that the methods are applicable to MIMO systems directly; and
\item that model fitting can improve with the introduction of multiple data sets, thereby unlocking the data archive.
\end{enumerate}
The specific gas turbine system and its identified models are not examined in detail since, at this stage, we are looking at the example as a proof of concept for the methodology.

Data sets were collected from an operating 10 megawatt dry low-emissions industrial gas turbine driving a centrifugal gas compressor in a natural gas pipeline application. These data sets are historical and were not artificially excited for system identification. In this case, data is sampled hourly and aggregated remotely at a central monitoring facility spanning a two month period. The gas turbine unit is periodically shut down for maintenance or due to a system fault. Each shut-down event creates a gap in the data record. Of the 17 contiguous data records, only 4 (records~5, 6, 15, 17) are of length close to 200 hours: 245, 265, 172 and 193 respectively, thereby limiting both the fidelity and model order achievable from a single data record. Each data set is 2-input, fuel flow $W_F$ and bleed valve command $BV_c$, and 2-output, Stage~5 temperature $T_5$ and shaft speed $N_{GP}$. Individual data sets 5 and 6 and then the multiple data sets $\{5,6\}$ and $\{5,6,15\}$ were used to fit  fourth-order models with
\begin{align}\label{eq:params}
\ell=5,\;\;m=2,\;\;n=4,\;\;p=2,
\end{align}
using the algorithm of Section~\ref{sec:mSSI}. 

\begin{figure}[ht]
\begin{center}
\includegraphics[width=85mm]{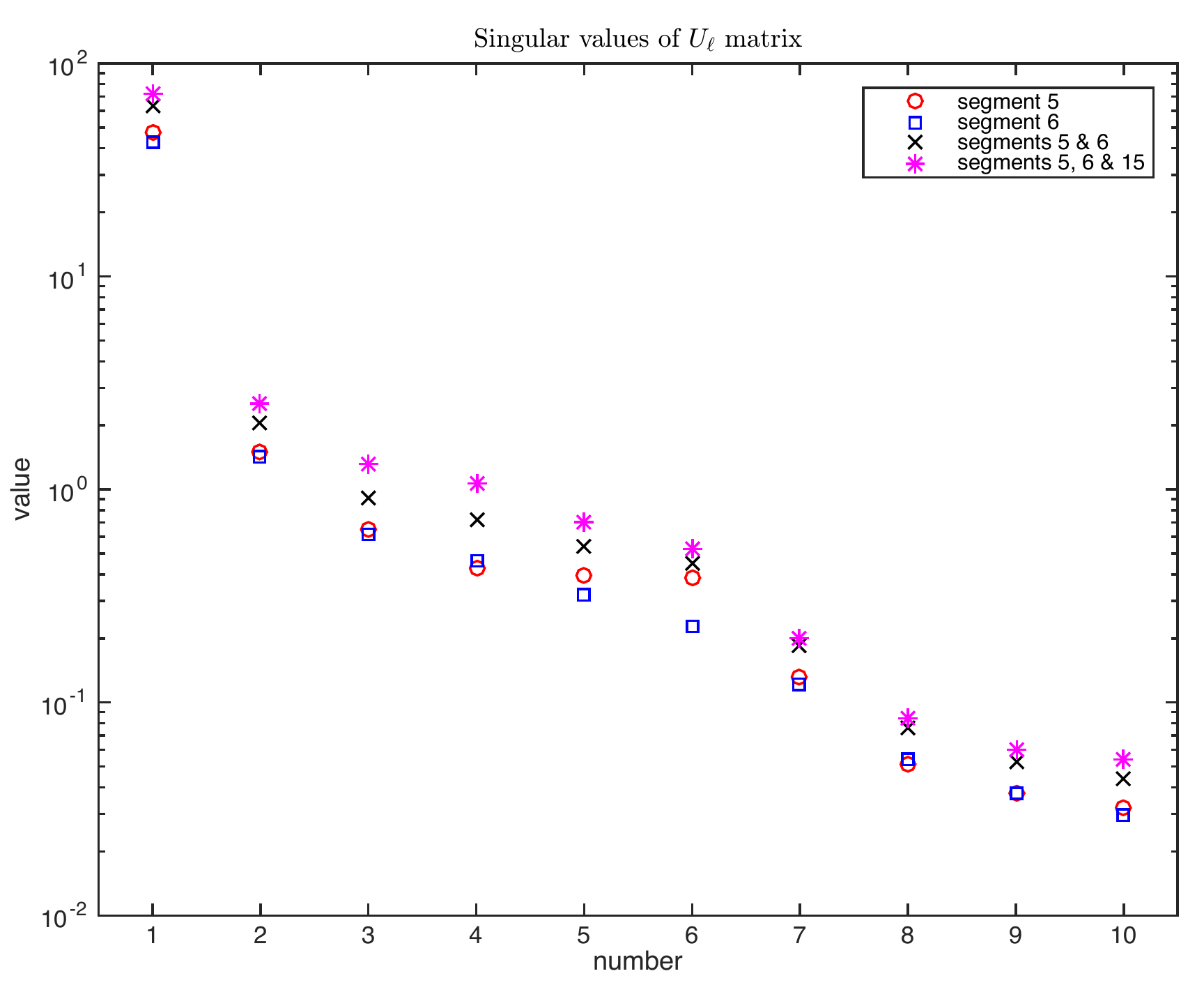}
\caption{Singular value distributions of the input matrix $\mathbb U_\ell$ from \eqref{eq:UUl} and Theorem~\ref{th:youBeauty}, which requires the rank to be 10.\label{fig:Usvd}}
\end{center} 
\end{figure}

\begin{figure}[ht]
\begin{center}
\includegraphics[width=85mm]{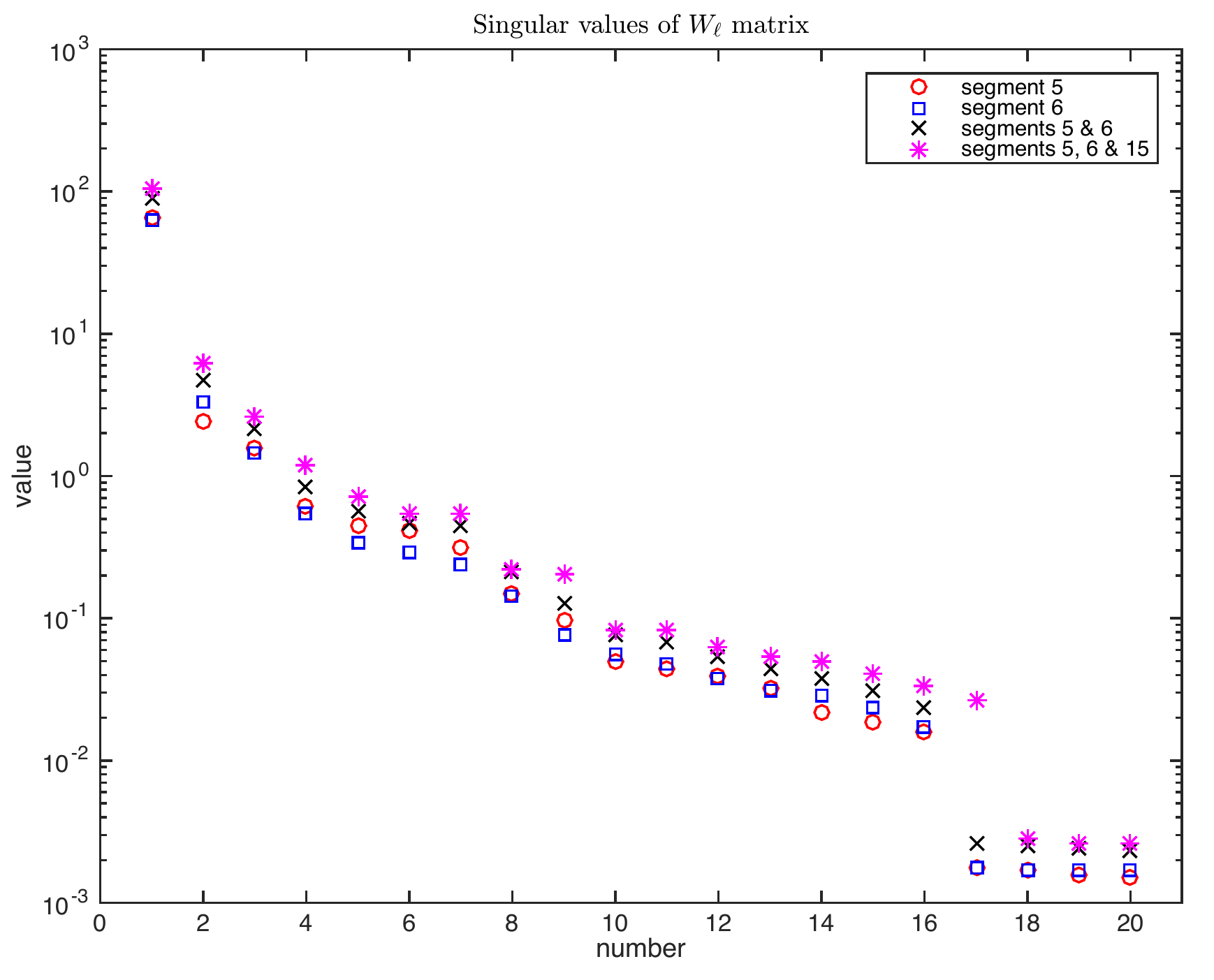}
\caption{Singular value distributions of the input-output matrix $\mathbb W_\ell$
from Theorem~\ref{th:youBeauty}, which requires the minimal rank to be 14.\label{fig:Wsvd}}
\end{center} 
\end{figure}

Figure~\ref{fig:Usvd} displays the singular values of the matrix $\mathbb U_\ell$ from \eqref{eq:UUl} while Figure~\ref{fig:Wsvd} shows those of the combined input-output matrix $\mathbb W_\ell=\begin{bmatrix}\mathbb U_\ell^T&\mathbb Y_\ell^T\end{bmatrix}^T$. The minimal rank condition for $\mathbb U_\ell$ from \eqref{eq:Urank} for the parameter values \eqref{eq:params} is $\ell m=10.$ The corresponding rank condition \eqref{eq:Wrank} for $\mathbb W_\ell$ is $\ell m+n=14.$ These two figures demonstrate the improvement in singular values achieved by incorporating further data sets into the identification; the joint data sets possess uniformly larger singular values than their constituents. 

The dimensions of the regressor matrices $\Upsilon$ from \eqref{eq:Upsilon} are: 490$\times$16 for the 245-data segment 5; 530$\times$16 for the 265-data segment 6; and 1020$\times$20 for the combined data segments 5 and 6; and 1364$\times$24 for the combined $\{5,6,15\}$ segments. With the number of rows given by number of input data (2 channels per time sample) and the number of columns given by the parameters: 8 for $4\times 2$ $B$, 4 for $2\times 2$ $D$, and 4 for each $4\times 1$ initial state. In each case, $\Upsilon$ is full rank computed by svd.  Although for the multiple segments~$\{5,6,15\}$ case, the identified $A$ has eigenvalues outside the unit circle, which leads to ill-conditioning of $\Upsilon$.

The performance of the models identified using data set~5, data set~6, and data sets~$\{5,6\}$ is evaluated using data set~17, an independent validation set of duration 193 hours. (The $\{5,6,15\}$ multi-segment model was not considered because of its instability.) The MATLAB function \texttt{compare} and data set 17 are used to evaluate the one-step-ahead prediction performance of the three models. The resulting mean squared prediction error performance for each output channel is presented in Table~\ref{tab:errf}.
\begin{table}[ht]
\centering 
\begin{tabular}{c c c c c c c} 
\toprule
RMS error ($\times 10^2$) &Model$_5$&Model$_6$&Model$_{\{56\}}$\\
$N_{GP}$ (output 1)&1.1266&1.0895&1.0796\\
$T_5$ (output 2)&1.0916&1.0980&0.9093\\
\bottomrule 
\end{tabular}
\caption{Root mean square (RMS) one-step-ahead prediction error from the three models with independent validation data set~17.\label{tab:errf}}
\end{table}

The conclusion to be reached from this analysis is that the capability to accommodate multiple MIMO data segments into the system identification process has a significant capacity to improve the quality of fit. In the specific gas turbine example, this effect is evident in the improved data quality measured by the singular values of the data matrices in Theorem~\ref{th:youBeauty} and largely due to the increased number of input-output data pairs. By the same token, had one or both data sets proven under-excited there is the possibility that the joint experiment might be sufficiently excited.

\section{Conclusion}

A new formal approach to the incorporation of multiple data sets into linear system identification has been developed and the behavioral theory of linear systems applied and extended to yield easily tested sufficient conditions on the collection of data sets for exact model identifiability. It is demonstrated that the accommodation of the additional initial conditions associated with non-contiguous data sets is achieved automatically by the method. The motivation for this study is the availability of significant quantities of industrial data in cloud-based archives of operational records, which are likely fragmented and discontinuous in time while also being largely under-excited. By advancing the prospect of system identification using multiple data sets and by providing a sufficiency test for identifiability, we are well placed to commence the profitable analysis of this archive. A proof-of-concept MIMO example using industrial gas turbine data was provided and demonstrated improved quantified model fitting performance.

Evidently, there remains considerable further work connected with the suite of options associated with subspace system identification, such as the choice of weighting matrices, instrumental variable steps, guaranteeing stability, etc. Likewise, on the data management side, approaches are needed to questions of automating the searching, sorting and selection of segments of data records to achieve sufficiently informative sets from which to fit reliably models for dognostics. The combination of industrial data archiving, multi-segment system identification, and the computer science of data management should represent an important foray into so-called \textit{Big Data}.

\bibliographystyle{plain}
\bibliography{/Users/bob/tex/bob}

\begin{thebibliography}{10}

\bibitem{CarrettBastinGeninGevers_TSP:1996}
P.~Carrette, G.~Bastin, Y.Y. Genin, and M.~Gevers.
\newblock Discarding data may help in system identification.
\newblock {\em IEEE Transactions on Signal Processing}, 44:2300--2310, 1996.

\bibitem{HolcombBitmead_TurboExpo:2015}
C.M. Holcomb and R.R. Bitmead.
\newblock Gas turbine analytics using archival data sets.
\newblock In {\em Proce ASME Turbo Expo: GT2015}, page to appear, Montreal,
  Canada, 2015.

\bibitem{LarimoreCDC:1990}
W.E. Larimore.
\newblock Canonical variate analysis in identification, filtering, and adaptive
  control.
\newblock In {\em Proceedings of the 29th {IEEE} Conference on Decision and
  Control}, volume~2, pages 596--604, 1990.

\bibitem{MarkovskyApproxCDC:2013}
I.~Markovsky.
\newblock Approximate system identification with missing data.
\newblock In {\em 52nd IEEE Conference on Decision and Control}, pages
  156--161, Florence Italy, 2013.

\bibitem{MarkovskyExactCDC:2013}
I.~Markovsky.
\newblock Exact system identification with missing data.
\newblock In {\em 52nd IEEE Conference on Decision and Control}, pages
  151--155, Florence Italy, 2013.

\bibitem{MarkovskyWillemsVanHuffelDeMoor:2006}
I.~Markovsky, J.C. Willems, S.~Van Huffel, and B.~De Moor.
\newblock {\em Exact and Approximate Modeling of Linear Systems: a Behavioral
  Approach}.
\newblock Mathematical Modeling and Computation. SIAM, Philadelphia PA, 2006.

\bibitem{VanOverscheeDeMoorAutom:1994}
P.~Van Overschee and B.~De Moor.
\newblock {N4SID}: {S}ubspace algorithms for the identification of combined
  deterministic-stochastic systems.
\newblock {\em Automatica}, 30(1):75--93, 1994.

\bibitem{PeretzkiIsakssonBittencourtForsman_AIChE2011}
D.~Peretzki, A.J. Isaksson, A.C. Bittencourt, and K.~Forsman.
\newblock Data mining of historic data for process identification.
\newblock In {\em Proceedings of AIChE Annual Meeting}, pages 1027--1033,
  Minneapolis MN, 2011.

\bibitem{VanOverschee&DeMoor:96}
Peter Van~Overschee and Bart De~Moor.
\newblock {\em Subspace identification for linear systems: theory,
  implementation, applications}.
\newblock Kluwer Academic Publishers, Dordrecht, 1996.

\bibitem{Verhaegen:94}
Michel Verhaegen.
\newblock Identification of the deterministic part of mmo state space models
  given in innovations form from input-output data.
\newblock {\em Automatica}, 30(1):61--74, Jan. 1994.

\bibitem{willems_persistency_2005}
J.~Willems, P.~Rapisarda, I.~Markovsky, and B.~De~Moor.
\newblock A note on persistency of excitation.
\newblock {\em Systems \&\ Control Letters}, 54(4):325--329, April 2005.

\bibitem{Willems_Autom:86}
Jan~C. Willems.
\newblock From time series to linear system --- {P}art {I}.. {F}inite
  dimensional linear time invariant systems.
\newblock {\em Automatica}, 22(5):561--580, 1986.

\bibitem{WillemsRapisardaMarkovskyDeMoorSCL:2005}
J.C. Willems, P.~Rapisarda, I.~Markovsky, and B.~De Moor.
\newblock A note on persistence of excitation.
\newblock {\em Systems \&\ Control Letters}, 54(4):325--329, 2005.

\bibitem{Wolovich:1974}
W.A. Wolovich.
\newblock {\em Linear Multivariable Systems}.
\newblock Springer-Verlag, New York, NY, 1974.

\bibitem{ZhangBitmeadTSP:2008}
Chengjin Zhang and R.R. Bitmead.
\newblock {MIMO} eqaulization with state-space channel models.
\newblock {\em IEEE Transactions on Signal Processing}, 56(10):5222--5231,
  2008.

\bibitem{ZhangBitmeadAutom:05}
Chengjin Zhang and R.R.Bitmead.
\newblock Subspace system identification for training-based {MIMO} channel
  estimation.
\newblock {\em Automatica}, 41:1623--1632, 2005.

\end{thebibliography}
\end{document}

\section{Optional pics with three-segment SysID}
Unfortunately, the model from $\{5,6,15\}$ is not stable, which gives simulation issues.
\begin{table}[ht]
\centering 
\begin{tabular}{c c c c c c c } 
\toprule
RMS error ($\times 10^2$) &Model$_5$&Model$_6$&Model$_{\{56\}}$&Model$_{\{5,6,15\}}$\\
$N_{GP}$ (output 1)&1.1349&1.0910&1.0888&2.3096\\
$T_5$ (output 2)&1.0437&1.1440&0.9510&8.2568\\
\bottomrule 
\end{tabular}
\caption{Root mean square (RMS) error between simulated outputs from the four models and measured outputs with input signals from independent validation data set~17.\label{tab:Eerrf}}
\end{table}

\begin{figure}[ht]
\begin{center}
\includegraphics[width=85mm]{UUsvd.pdf}
\caption{Singular value distribution of the input matrix $\mathbb U_\ell$ from \eqref{eq:UUl} and Theorem~\ref{th:youBeauty}. The minimal rank required of this matrix is 10 according to the theorem.\label{fig:UUsvd}}
\end{center} 
\end{figure}
\begin{figure}[ht]
\begin{center}
\includegraphics[width=85mm]{WWsvd.pdf}
\caption{Singular value distribution of the input-output matrix $\mathbb W_\ell$
from Theorem~\ref{th:youBeauty}. The minimal rank required of this matrix is 14 according to the theorem.\label{fig:WWsvd}}
\end{center} 
\end{figure}

\end{document}